\documentclass[prl,twocolumn,amsmath,amssymb,superscriptaddress]{revtex4}
\usepackage{graphicx}
\usepackage{dcolumn}
\usepackage{bm}

\begin{document}
\preprint{cond-mat/999999}

\title{ Persistence of Strong Electron Coupling to a 
Narrow Boson Spectrum in Overdoped Bi$_{2}$Sr$_{2}$CaCu$_{2}$O$_{8+ \delta}$
Tunneling Data}


\author{J.F. Zasadzinski$^{1,2}$, L. Ozyuzer$^{2,3}$ L. Coffey$^{1}$, 
K.E. Gray$^{2}$, D.G. Hinks$^{2}$ and C. Kendziora$^{4}$}

\affiliation{1.Illinois Institute of Technology, Chicago, IL 60616\\
2.Materials Science Division, Argonne National Lab, Argonne IL 60439\\
3.Izmir Institute of Technology, TR-35437 Izmir, Turkey\\
4.Naval Research Laboratory, Washington DC 20375}

\date{\today}

\begin{abstract}
A d-wave, Eliashberg analysis of break junction and STM tunneling 
spectra on   Bi$_{2}$Sr$_{2}$CaCu$_{2}$O$_{8+ \delta}$
(Bi2212) reveals that the spectral dip feature is directly linked 
to strong electronic coupling to a narrow boson spectrum,
evidenced by a large peak in $\alpha{^2}F(\omega)$.  The tunneling dip 
feature remains robust in the overdoped regime of Bi2212 with 
bulk T$_{c}$ values of 56 K-62 K.  This is contrary to recent 
optical conductivity measurements of the self-energy that 
suggest the narrow boson spectrum disappears in overdoped 
Bi2212 and therefore cannot be essential for the pairing mechanism.  
The discrepancy is resolved by considering the way each technique 
probes the electron self-energy, in particular, 
the unique sensitivity of tunneling to the off-diagonal 
or pairing part of the self-energy.
\end{abstract}


\maketitle 
 
Recently, an important consensus has been reached among various 
spectroscopies which probe electron interactions 
in high T$_{c}$ superconductors (HTS).  For nearly optimal doped 
Bi$_{2}$Sr$_{2}$CaCu$_{2}$O$_{8+ \delta}$ (Bi2212), tunneling\cite{1,2}, 
angle-resolved photoemission (ARPES)\cite{3,4,5} and the 
Drude part of the optical conductivity\cite{6,7} have all exhibited spectral 
evidence that the electrons which participate in superconductivity 
are coupled to a relatively narrow boson spectrum peaked at an energy of 35-43 meV.  
The nature of this boson mode[8] and its relevance to the high T$_{c}$
mechanism remain under intense debate.  Strong-coupling analyses of 
Bi2212 tunneling\cite{9} and related optical 
conductivity in YBa$_{2}$Cu$_{3}$O$_{7}$ \cite{10} have shown that coupling 
to this mode alone is sufficient to explain high T$_{c}$ 
superconductivity.  However, in a doping-dependent optical conductivity study of 
Bi2212\cite{7}, the mode's importance has been downplayed since it seems to 
disappear into a broad background of excitations in the 
overdoped regime where T$_{c}$ is still high, $ \simeq$ 60 K.   To address this 
important issue we examine previously published break junction 
tunneling measurements on Bi2212, and some new data, that includes heavily overdoped crystals
with T$_{c}$ values of 56 K-62 K.   A quantitative Eliashberg analysis shows 
that the tunneling spectral dip feature is directly linked to a relatively 
narrow, dominant peak in the electron-boson spectral function, 
$\alpha^{2} F(\omega)$.  Tunneling data from different overdoped crystals show 
reproducibly that the dip (and therefore coupling to the mode) remains a robust feature, showing no 
evidence of disapppearing.   This apparent contradiction with optical 
conductivity may be explained by considering the unique sensitivity 
of tunneling to the off-diagonal, or pairing part the electronic self-energy.   

Each of the above electron spectroscopies reveals (within specific 
experimental limitations) the complex quasiparticle self-energy, 
a matrix which contains all information on electron interactions 
including, presumably, those responsible for superconductivity.  
In addition to the diagonal part, $\Sigma(\omega)$, 
a superconductor has an off-diagonal part, $\phi(\omega)$, due to 
the electron-paired condensate.   Peaks in the real 
part of the optical self energy, $- Re\Sigma(\omega)$, \cite{6,7,10}, 
kinks and dips in ARPES \cite{3,4,5}, and dips in tunneling 
conductance \cite{1,2} are all consistent with electrons coupling to 
a bosonic mode in HTS.  The mode has been argued to be 
the resonance spin excitation \cite{1,2,3,4,6,7} found in neutron scattering,
\cite{11} the B1g optical phonon,\cite{5} or perhaps a magnetic 
polaron.\cite{6} One of the main causes of ambiguity in the interpretation 
of electron spectroscopies is that optimal doped Bi2212 presents a conspiracy 
of similar values for the energies of interest.  
The superconducting gap parameter, $\Delta$, the B1g phonon and 
the resonance spin excitation all have energies 
in the range 35 meV - 43 meV.  On the other hand these quantities 
have distinct doping dependencies. Raman spectroscopy \cite{12} has shown 
that the B1g phonon is essentially independent of hole concentration, 
remaining at $\simeq$ 35 meV , whereas the resonance spin excitation 
is proportional to T$_{c}$.\cite{11}   To extract the mode energy from any spectroscopy 
requires a realistic model and it is desirable that the experiment provide a 
direct measure of $\Delta$ (as does tunneling) since this quantity also 
enters the quasiparticle self-energy and is known to have a strong 
doping dependence. \cite{13} 
  
Here we present quantitative fits of published \cite{1,13,14,15} superconductor-insulator-superconductor 
(SIS) break-junction tunneling conductances and an SIN (N = normal metal) 
conductance obtained by scanning tunneling microscopy (STM), 
using a self-consistent Eliashberg theory and an electron-boson spectral 
function $\alpha^{2}F(\omega)$.  
For near optimal doped Bi2212 the resulting $Re \Sigma(\omega)$
 bears a strong resemblance to that extracted directly from 
ARPES data \cite{4} thereby linking two different electron spectroscopies.  
Combining a new SIS tunneling conductance with published data on overdoped Bi2212 
with $\Delta$ values in the range of 17 meV - 19 meV,  a robust and highly 
reproducible dip feature is established.  This doping range  corresponds 
to T$_{c}$ $\simeq$ 60 K, which is the value where 
the mode supposedly disappears in optical conductivity.  A fit of the 
most overdoped Bi2212 SIS conductance demonstrates how $\alpha^{2}F(\omega)$ 
changes in going from near optimal to heavily overdoped and the tunneling
$\Sigma(\omega)$ gives 
important insights into the discrepancies with optical conductivity.\cite{7}  

The analysis begins with a simultaneous, quantitative fit of a break 
junction SIS conductance\cite{1} and an SIN conductance\cite{9} obtained by STM.  
States-conserving normalization was accomplished by fitting the high bias conductance to a smooth 
polynomial as described in ref. \cite{9} 
A self-consistent, d-wave Eliashberg formalism \cite{9} was used to generate 
the quasiparticle density of states (DOS).  Both data sets were 
obtained on slightly overdoped Bi2212 
and thus a similar $\alpha^{2}F(\omega)$ would be expected if indeed 
the dip feature were a strong-coupling effect. 
The data and fits are shown in Fig. 1 and the 
corresponding $\alpha^{2}F(\omega)$ is shown as \#1 in 
Fig. 3(a).  The STM data set (Fig. 1 inset) was analyzed previously using 
the same procedure \cite{9} but 
with an $\alpha^{2}F(\omega)$ consisting only of a narrow, 
Lorentzian boson spectrum which was adjusted to best fit 
the data.   In this analysis the $\alpha^{2}F(\omega)$ \#1 includes a 
broad, higher energy spectrum out to 160 meV in 
addition to the sharp mode peaked at 39 meV as suggested in both ARPES \cite{4} 
and optical conductivity. \cite{7} The DOS fit to the normalized STM data 
is improved by including this higher energy tail in  $\alpha^{2}F(\omega)$ in that 
the dip strength is more closely matched.  The fit also captures the shoulder, 
or strong coupling onset feature, in the STM data, which is due to the 
low frequency threshold in $\alpha^{2}F(\omega)$ \#1 near 17 meV.

The same $\alpha^{2}F(\omega)$ \#1 provides a good fit of the 
SIS break junction conductance \cite{1} in Fig. 1.  
A particular break junction was chosen which had a similar gap value 
as the STM data but also had a peak height to background ratio (PHB) 
which was not too large ($\simeq$ 3) for reasons which will become clear shortly. 
To accommodate the slight differences in $\Delta$ values the SIS data and fit are 
plotted on a rescaled voltage axis 
which is in units of $\Delta$.   While further fine-tuning of $\alpha^{2}F(\omega)$
could be done, the fits in Fig. 1 prove sufficiently that the dip structure 
is reproducible among different junction types and can be treated 
{\em quantitatively} as a strong coupling effect.  The shape and 
strength of the dip determines, self-consistently, the measured gap parameter,
$\Delta$.

\begin{figure}
\includegraphics{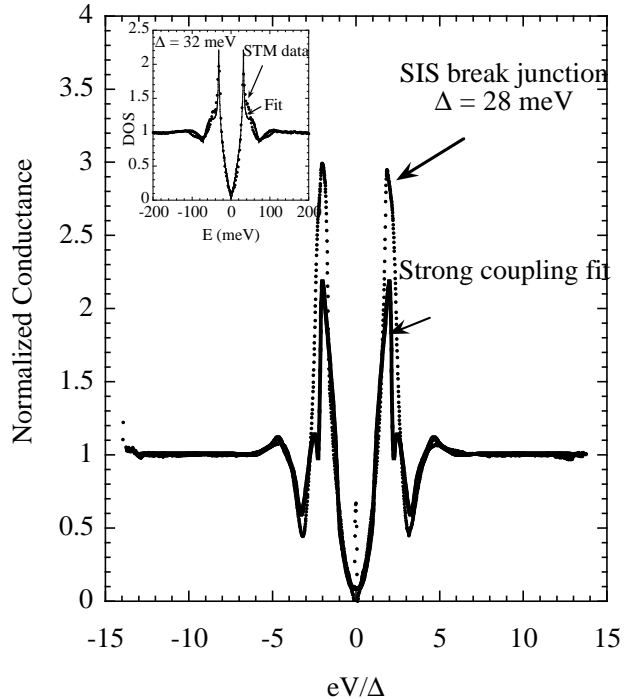}
\vspace{-6.25in}
\caption{ Comparison of normalized SIS break junction tunneling conductance 
(dots) and  d-wave Eliashberg fit (solid line) for a 
junction with $\Delta$ = 28 meV.  Inset: The same 
$\alpha^{2}F(\omega))$ is used to fit an SIN conductance obtained by STM.\\
 }
\end{figure}

Note the enhancement in the dip strength for the SIS calculation 
and experiment compared to the SIN result which is due to the convolution 
of the two DOS in the SIS conductance.\cite{16} This points out a 
particular advantage of SIS tunneling for probing electron interactions, 
but at a cost of not knowing the particle-hole symmetry of the DOS.  
However, the consistency of the SIN and SIS fits of Fig. 1 suggests that 
the symmetric 
dip features of the present STM data are intrinsic to the DOS and the common observation of asymmetries 
in the dip strength, e.g. in point-contact SIN tunneling, \cite{13} likely originate from 
some other effect.  Contributions to the tunnel conductance from asymmetric 
pseudogaps near defects, which exhibit a broad peak at positive bias, may be the cause as discussed in ref.\cite{9}.
     
An obvious difference between data and theory is in the PHB ratio
which is typically $\simeq$ 2 for SIS calculations which use a d-wave DOS 
as in Fig. 1 inset. We have previously pointed out \cite{1,14,15} that 
measured SIS break junctions often display much larger 
PHB ratios and that these most likely arise from the tunneling matrix 
element which favors tunneling 
along the ($\pi$,0) momentum or the anti-node directions.  
In our Eliashberg model, the complex gap 
parameter $\Delta(\omega, \phi)$ = $\Delta(\omega)$ cos(2$\phi$) where $\phi$ 
is the polar angle in momentum space.\cite{9}  Preferential 
tunneling along the antinode ($\phi$=0) will thus increase the dip 
strength since this is the direction of maximum amplitude of $\Delta(\omega, \phi)$.
The inclusion of a tunneling weighting factor in the 
calculation of the SIS conductance in Fig. 1, will increase both 
the PHB ratio and the dip strength 
which will improve the fit to the data.   However, here we chose not 
to include this additional parameter in the SIS fit so that the 
bare dip strength from the Eliashberg model could be observed.  
It is evident from Fig. 1 that preferential tunneling is not
necessary for the observation of strong dip features. 

The model also shows that the energy of 
the dip minimum relative to the gap edge provides a good estimate of the energy of the sharp mode, $\Omega$, in $\alpha^{2}F(\omega)$.   
The STM data fit leads to $\Delta$ = 32 meV  and $\Omega$ = 39 meV while the 
break junction in Fig. 1 has the values $\Delta$ = 28 meV and $\Omega$ = 34 meV. 
While the dominant feature in $\alpha^{2}F(\omega)$ is the sharp boson 
mode, and this predominantly determines the dip strength,
the inclusion of a broad, higher energy spectrum improves the fit 
and suggests that electron coupling to these excitations is also relevant 
to understanding superconductivity. 

The detailed, quantitative analysis described above provides an 
understanding of the tunneling 
conductances observed in heavily overdoped Bi2212.  Three normalized break 
junction tunneling conductances, both published\cite{1,14} 
and unpublished, from three different overdoped crystals (T$_{c}$ = 56-62 K) are 
shown in the inset of Fig. 2. Junctions have been chosen which have $\Delta$ values in the range 17 meV - 19 meV to examine 
the reproducibility of dip features for a given value of $\Delta$.  What is evident is that the three junctions 
exhibit a high degree of reproducibility in the shape, 
strength and characteristic voltage of the dip feature. Based
on the similarity of the overall conductance shapes to the SIS data analyzed 
in Fig. 1, it can be inferred directly that $\alpha^{2}F(\omega)$ 
will have a sharp peak characterized by an energy $\Omega \simeq$ 30 meV, 
obtained from the dip minimum position. Temperature dependent measurements 
\cite{15} lead to an estimated T$_{c}$$\simeq$ 60 K for these junctions, 
which is the value for which optical conductivity data \cite{7} 
indicates the disappearance of the sharp boson mode in overdoped Bi2212.  
The tunneling data show no evidence of such disappearance.

\begin{figure}
\includegraphics{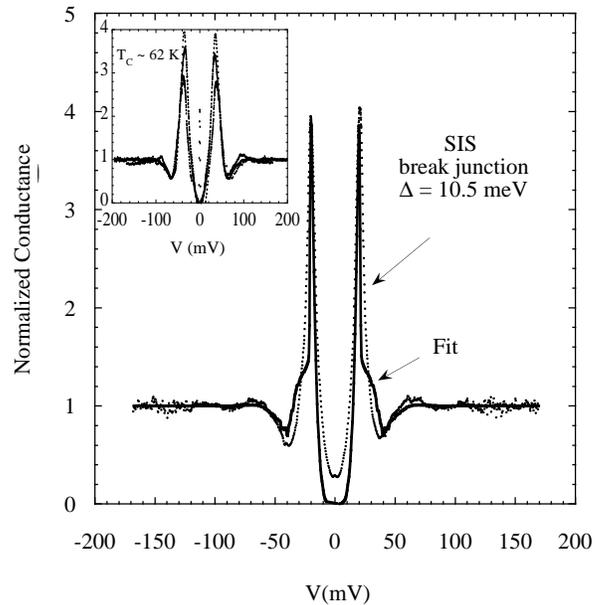}
\vspace{-6.75in}
\caption{  Comparison of normalized SIS break junction tunneling conductance (dots) 
and d-wave Eliashberg fit (solid line) for a junction on 
overdoped Bi2212 with $\Delta$ = 10.5 meV.  
Inset: A set of normalized break junction tunneling conductances on 
three different overdoped Bi2212 crystals.  Junctions have  been
chosen which have $\Delta$ values in the range 17 meV - 19 meV.  
The estimated T$_{c}$ based on temperature dependent measurements is ~ 60 K.\\
 }
\end{figure}

The main panel of Fig. 2 shows normalized published data and strong-coupling fit for a 
very heavily overdoped Bi2212 junction 
which has a measured junction T$_{c}$ $\simeq$ 56 K. \cite{15}  This junction 
exhibits a gap parameter $\Delta$ = 10.5 meV, which to our 
knowledge is the smallest value reported for Bi2212.  The dip 
features are adequately fit by the Eliashberg model, 
and this leads to the $\alpha^{2}F(\omega)$ \#2  
shown in Fig. 3(a).  Again, self-consistency is obtained 
as  $\alpha^{2}F(\omega)$ \#2 leads 
directly to the measured gap.  Here we have used a tunneling weighting 
factor (see ref. 15) to achieve the PHB ratio. The $\alpha^{2}F(\omega)$  
\#2 demonstrates that even in this very heavily overdoped 
crystal, the electronic coupling to the boson mode has not disappeared 
and in fact dominates the spectral function.

The principal issue raised by this study is why such a discrepancy exists 
between tunneling and optical conductivity spectra for overdoped Bi2212.   
Both experiments probe the entire Fermi surface so arguments 
based on momentum selectivity (e.g. ARPES [18]) do not apply.   
To compare with ARPES and optical conductivity, 
we plot -2$Re\Sigma(\omega)$ for the two $\alpha^{2}F(\omega)$  spectra 
in Fig. 3(b).  The curve \#1 corresponds to $\alpha^{2}F(\omega)$ \#1
and bears a remarkable resemblance to -2$Re\Sigma(\omega)$ determined 
directly from the low temperature nodal quasiparticle spectral weight 
in ARPES.\cite{4}  Furthermore, the peak has an amplitude near 500 cm$^{-1}$ 
which corresponds to the {\em difference} between 
superconducting and normal -2$Re\Sigma(\omega)$ in optical conductivity.\cite{7}
This implies that electron interactions are 
showing up in the optical conductivity that are not seen in the tunneling spectra. 
  
We believe the reason for the discrepancy lies in the way 
tunneling probes the electron self-energy.  
In conventional, s-wave superconductors, the tunneling DOS
$\simeq$ $1+1/2Re(\Delta(\omega)/\omega)^{2}$ where $\Delta(\omega)=
\phi(\omega)/Z(\omega)$ and the pairing self-energy, $\phi(\omega)$,   
primarily gives rise to the phonon fine structure.\cite{19,20}  
Above Tc, $\phi(\omega)=0$ and the tunneling DOS is flat and 
featureless (as found in experiment). Also beyond a cutoff frequency, 
$\phi(\omega)=0$, effectively decoupling 
superconductivity (and tunneling spectra) from higher frequency electron interactions.  
The d-wave, Eliashberg model\cite{9} used here has an additional momentum 
dependence, but the sensitivity to the pairing self-energy remains.  Therefore 
the tunneling $\Sigma(\omega)$, generated along with $\phi(\omega)$, contains 
only those electron interactions which participate in pairing. 

\begin{figure}
\includegraphics{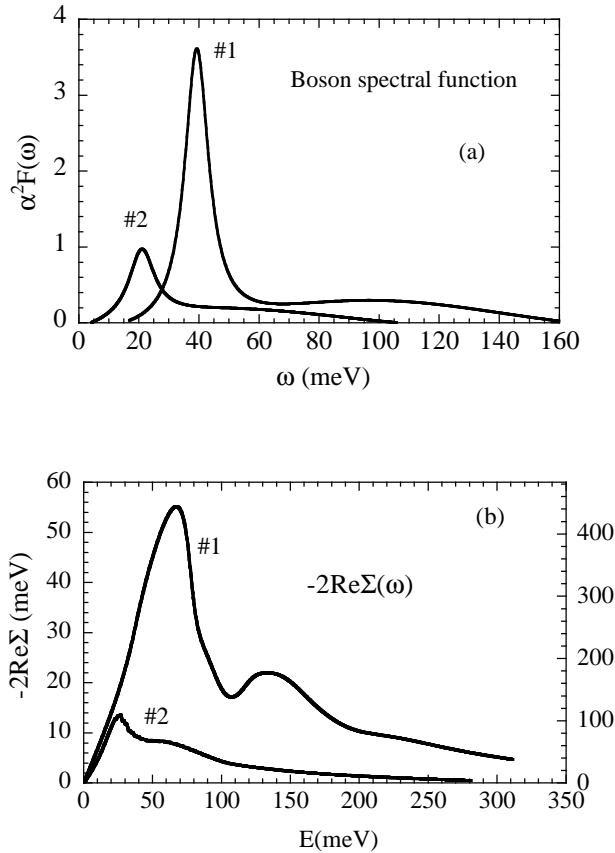}
\vspace{-5.5in}
\caption{ (a) Electron-boson functions,$\alpha^{2}F(\omega))$ , which result 
from the strong-coupling fits 
to the near optimal SIN data of Fig. 1 (inset) and the overdoped 
SIS data of Fig. 2, labeled 
as \#1 and \#2 respectively.  
(b) corresponding real part of the diagonal self-energies, -2$Re(\Sigma(\omega))$, 
obtained from the $\alpha^{2}F(\omega))$ shown in (a).     
 }
\end{figure}

On the other hand, optical conductivity probes the scattering 
rate and will reflect all electron 
interactions which enter the full diagonal 
self-energy $\Sigma(\omega)$.  The large size of 
the broad, high energy background relative to the resonance 
mode observed in optical self-energy is not 
compatible with the $\alpha^{2}F(\omega)$ or $\Sigma(\omega)$ found in Fig. 3.  
This implies that a large fraction of the high 
energy boson continuum indeed couples to electrons but is not relevant 
to superconductivity.  This is reminiscent of conventional superconductors 
where the high frequency part of the coulomb interaction 
plays no role in the superconductivity and this repulsive interaction 
enters $\phi(\omega)$ as $\mu^{*} \simeq 0.1$, reduced 
from the total electron-electron coupling constant,$\mu \simeq 1.0$ .
\cite{19,20} Thus the tunneling data indicate 
that the mode has not disappeared in the optical conductivity. 
Rather, Fig. 3(b) shows that (-2 $Re \Sigma(\omega))$ \# 2 
from tunneling is considerably reduced in size compared with \# 1 
and the mode becomes unresolved in a 
broad spectrum of excitations which do not participate in superconductivity. 

In summary, SIS break junction tunneling data on near optimal and 
heavily overdoped Bi2212 have been 
analyzed quantitatively to provide the electron-boson spectral function 
$\alpha^{2}F(\omega)$ and the diagonal self-energy $\Sigma(\omega)$.  
The robust dip feature is directly linked to strong electronic coupling 
to a narrow boson spectrum, a peak in $\alpha^{2}F(\omega)$,
which drives the superconductivity and shows no evidence of 
disappearing with overdoping. This fundamental disagreement with 
optical conductivity can be resolved by considering the way 
each experiment probes the electron self-energy.  Tunneling directly 
measures the pairing part,$\phi(\omega)$, 
and the resulting $\alpha^{2}F(\omega)$ and $\Sigma(\omega)$ reflect only 
those electronic interactions which participate in 
pairing.  The more detailed analysis presented here confirms previous reports\cite{1} 
that the mode energy $\Omega$ decreases substantially with overdoping which 
seems to rule out the B1g phonon and favors the resonance 
spin excitation as its origin.

Acknowledgements
This work was supported in part
\hspace*{0.1in}
by US Department of Energy, 
Division of Basic Energy 
\hspace*{0.1in} Sciences - 
Materials Science under contract no. W-31-
\hspace*{0.1in}   109-ENG-38.


\begin{thebibliography}{28}
\bibitem{1}J.F Zasadzinski et al, Phys. Rev. Lett. 87, 067005 (2001)
\bibitem{2}B. W. Hoogenboom, C. Berthod, M. Peter, O. Fischer, A.A. Kordyuk, 
Phys. Rev. B 67, 224502 (2003)
\bibitem{3}J.C. Campuzano et al, Phys. Rev. Lett. 83, 3709 (1999)
\bibitem{4}P.D. Johnson, T. Valla, A.V. Fedorov, Z. Yusof, B. O. Wells, Q. Li, 
A. R. Moodenbaugh, 
G. D. Gu, N. Koshizuka, C. Kendziora, Sha Jian and D. G. Hinks, 
Phys. Rev. Lett. 87, 177077 (2001)
\bibitem{5}T. Cuk et al, (unpublished) cond-mat 0403521
\bibitem{6}J.J. Tu, C.C. Homes, G.D. Gu, D.N. Basov and M. Strongin, 
Phys. Rev. B 66, 144514 (2002)
\bibitem{7}J. Hwang, T. Timusk, G. D. Gu, Nature 427, 714 (2004)
\bibitem{8}For simplictiy we use the term boson mode and narrow 
spectrum of excitations interchangeably
\bibitem{9}J.F. Zasadzinski, L. Coffey, P. Romano, Z. Yusof,
 Phys. Rev. B 68, 180504, (2003)
\bibitem{10}J. P. Carbotte, E. Schachinger and D.N. Basov, Nature (London) 401, 354 (1999)
\bibitem{11}For a review of the resonance mode see Y. Sidis, 
S. Pailhes, B. Keimer, P. Bourges, C. Ulrich 
and L.P. Regnault, Phys. Stat. Sol (2003)
\bibitem{12}C. Kendziora and A. Rosenberg, Phys. Rev. B 52, 9867 (1995)
\bibitem{13}N. Miyakawa, J.F. Zasadzinski, L. Ozyuzer, P. Guptasarma, 
D. G. Hinks, C. Kendziora, K.E. Gray, Phys. Rev. Lett. 83, 1018 (1999)
\bibitem{14}L. Ozyuzer, J.F. Zasadzinski, C. Kendziora, K.E. Gray, 
Phys. Rev. B61, 3629 (2000)
\bibitem{15}L. Ozyuzer, J.F. Zasadzinski, K.E. Gray, C. Kendziora, N. Miyakawa, 
Europhys. Lett. 58, 589 (2002)
\bibitem{16}The SIS fit also shows an additional, sharper dip feature between 
the peak and dip which arises from the strong coupling onset in the DOS.  
While no such sharp structure has ever been observed 
in the break junction conductances, it is not uncommon to 
observe a shoulder in SIS conductances just beyond the peak, which might be 
related to this effect.  It must be remembered that these 
break junctions are macroscopic and any inhomogeneities will broaden such 
sharp features.
\bibitem{17}Lack of agreement in the sub-gap region may be due in part 
to an incipient Josephson current.  
Well-defined Josephson currents are typically observed in moderately 
overdoped SIS break junctions.
\bibitem{18}T. Cuk, A.D. Gromko, Zhe Sun, Z.-X. Shen, D.S. Dessau (unpublished) 
cond-mat 0403743
\bibitem{19}E. L. Wolf, {\em Principals of Electron Tunneling Spectroscopy} 
(Oxford University Press, Oxford, 1985)
\bibitem{20}W. L. McMillan and J.M. Rowell, in {\em Superconductivity} 
edited by R.D. Parks, (
Dekker, New York, 1969) 

\end{thebibliography}
\end{document}